\documentclass[twocolumn,preprintnumbers,amsmath,amssymb,superscriptaddress]{revtex4}
\usepackage{graphicx}
\usepackage{dcolumn}
\usepackage{bm}
\usepackage{soul}
\usepackage{color}
\usepackage{epstopdf}
\usepackage[version=3]{mhchem}
\usepackage{lipsum}
\usepackage[outercaption]{sidecap}
\usepackage{floatrow}
\usepackage{hyperref}
\usepackage{multirow}

\begin{document}

\title{Multiscale Prediction of Polymer Relaxation Dynamics via Computational and Data-Driven Methods}

\author{Nguyen T. T. Duyen}
\affiliation{Faculty of Materials Science and Engineering, Phenikaa University, Hanoi 12116, Vietnam}
\author{Ngo T. Que}
\affiliation{Phenikaa Institute for Advanced Study, Phenikaa University, Hanoi 12116, Vietnam}
\author{Anh D. Phan}
\email{anh.phanduc@phenikaa-uni.edu.vn; Present address: Center for Materials Innovation and Technology, VinUniversity, Hanoi, Vietnam.}
\affiliation{Faculty of Materials Science and Engineering, Phenikaa University, Hanoi 12116, Vietnam}
\affiliation{Phenikaa Institute for Advanced Study, Phenikaa University, Hanoi 12116, Vietnam}
\date{\today}
\begin{abstract}
We present a multiscale modeling approach that integrates molecular dynamics simulations, machine learning, and the Elastically Collective Nonlinear Langevin Equation (ECNLE) theory to investigate the glass transition dynamics of polymer systems. The glass transition temperatures ($T_g$) of four representative polymers are estimated using simulations and machine learning model trained on experimental datasets. These predicted $T_g$ values are used as inputs to the ECNLE theory to compute the temperature dependence of structural relaxation times and diffusion coefficients, and the dynamic fragility. The $T_g$ values predicted from simulations show good quantitative agreement with experimental data. While machine learning tends to slightly overestimate $T_g$, the resulting dynamic fragility values remain close to experimental fragilities. Overall, ECNLE calculations using these inputs agree well with broadband dielectric spectroscopy results. Our integrated approach provides a practical and scalable tool for predicting the dynamic behavior of polymers, particularly in systems where experimental data are limited.
\end{abstract}

\maketitle
\section{Introduction}
Organic polymers such as poly(phenylene sulfide) (PPS), poly(isoprene) (PI), poly(propylene glycol) (PPG), and poly(butadiene) (PB) have been widely studied due to their valuable thermal, mechanical, and chemical properties. PI and PB are elastomeric polymers with double bonds in their backbone, which provide high chain flexibility. This makes them suitable for applications requiring elasticity such as tires, seals, and conveyor belts \cite{3,5}, as well as biomedical devices\cite{2,7}. PPG is a flexible polymer capable of forming hydrogen bonds and shows good chemical stability across a range of environments. It is commonly used in coatings, adhesives, and biocompatible products \cite{26,27,28,29}. In contrast, PPS has a rigid backbone containing sulfur atoms. Since this structural feature improves its thermal stability and chemical resistance, PPS can be used under demanding engineering conditions \cite{8}. To improve the performance of such polymers, an accurate determination and control of their glass transition temperature and related properties is essential. Despite extensive research, the fundamental nature of the glass transition remains a central challenge in the study of amorphous materials. 

Several experimental techniques are commonly used to determine the glass transition temperature ($T_g$). including differential scanning calorimetry (DSC) \cite{13}, dynamic mechanical analysis (DMA) \cite{14,15}, and broadband dielectric spectroscopy (BDS) \cite{16}. Each method has distinct advantages and disadvantages. DSC has been widely adopted because of its simplicity and ability to detect heat flow associated with thermal transitions. However, when the glass transition overlaps with melting or crystallization events, it is hard to determine $T_g$. DMA has been widely used to measure the temperature dependence of viscoelastic behaviors of polymers with significant mechanical changes typically observed near $T_g$. However, it requires precise preparation of sample dimensions and careful mechanical calibration to ensure accurate results. BDS data provide detailed insight into molecular relaxation dynamics over a broad frequency range. It allows us to calculate the temperature and pressure dependence of structural relaxation times, $T_g$, and the dynamic fragility. Despite its capabilities, BDS requires specialized instrumentation, can be costly, and is not universally applicable to all types of materials. These limitations have led to the development of computational approaches that include molecular dynamics (MD) simulations and theoretical models to complement or substitute the experimental determination of $T_g$ and structural relaxation time.

MD simulations provide atomistic insight into the structural and thermal behavior of polymers by tracking the time evolution of particles under realistic interatomic interactions. The glass transition temperature can be estimated from such simulations by analyzing the temperature dependence of volume or density. However, MD simulations are limited to short timescales, typically up to $10^5$ picoseconds. Meanwhile, experimental observations often span milliseconds to hundreds of seconds. It means that MD simulations cannot fully capture the slow relaxation dynamics characteristic of systems near and below $T_g$.

To address the timescale limitations of MD simulations, the Elastically Collective Nonlinear Langevin Equation theory has been developed to describe structural relaxation in glass-forming systems. The ECNLE theory describes how local cage-scale hopping cooperates with long-range elastic displacements to govern structural relaxation. It predicts the temperature dependence of relaxation times and diffusion coefficients across a wide dynamic range from picoseconds to hundreds of seconds, which cover both simulation and experimental timescale. This theory has been successfully applied to various amorphous materials including polymers \cite{19,20}, molecular liquids, organic compounds \cite{21,22}, metallic glasses \cite{23}, and confined systems. ECNLE predictions have shown strong agreement with experimental data and/or simulation \cite{19,20,21,22,23}. A key input to ECNLE calculations is $T_g$, which provides the thermal mapping from density to temperature. However, for newly synthesized or computationally designed materials, experimental $T_g$ values are often unavailable and this restricts the applicability of the theory.

To overcome this challenge, in this study, we present an integrated approach that combines MD simulations, machine learning, and ECNLE theory to investigate the glassy dynamics of representative polymer systems. First, MD simulations and trained ML models are used to estimate the $T_g$ values of each polymer system based on structural or compositional information. These predicted $T_g$ values and experimental counterparts are then used as inputs for the ECNLE theory through the thermal mapping. We subsequently compute the temperature dependence of the structural relaxation time and diffusion coefficient, and the dynamic fragility. Finally, we validate our predictions by comparing them against available experimental data. This integrated MD–ML–ECNLE approach provides a predictive and scalable platform for understanding and designing amorphous materials, particularly when experimental data are limited or unavailable.

\section{THEORETICAL BACKGROUND}
In this section, we present the theoretical and computational methods used in this study. We first describe how the ECNLE theory is applied to predict the temperature dependence of structural relaxation dynamics. We then introduce MD simulations and ML models for estimating $T_g$ of polymers. By integrating these techniques, we establish a multiscale framework capable of predicting both $T_g$ and relaxation behavior for polymer systems without relying on adjustable or experimental parameters. Each approach is described in the following subsections.
\subsection{The ECNLE theory}
In the ECNLE theory, activated events in amorphous systems are analyzed by describing the material as a dense hard-sphere fluid system \cite{19,20,21,22,23,30,31,32,33,34,35,ECNLE1, ECNLE2, 36,37,38}. The model treats particles as hard spheres with a number density, $\rho$, and a characteristic diameter, $d$. The dynamics of a particle are influenced by short-range interactions with neighboring particles, thermal noise, and frictional damping. These effects are collectively described by a nonlinear Langevin equation. Solving this equation provides the dynamic free energy, $F_{\text{dyn}}(r)$, which quantifies the effective confinement or caging constraints imposed on the tagged particle by its surroundings \cite{19,20,21,22,23,30,31,32,33,34,35,ECNLE1, ECNLE2, 36,37,38}
\begin{eqnarray} 
\frac{F_{\text{dyn}}(r)}{k_B T} &=& -3\ln \left(\frac{r}{d}\right) \nonumber\\
&-& \int_0^\infty \frac{q^2 d^3 [S(q)-1]^2}{12\pi \Phi [1 + S(q)]} \exp\left[-\frac{q^2 r^2 (1 + S(q))}{6 S(q)}\right] dq,\nonumber\\
\label{eq:1}
\end{eqnarray}
where $k_B$ is the Boltzmann constant, $T$ is the ambient temperature, $r$ is the displacement, and $q$ is the wavevector, $S(q)$ is the static structure factor which can be obtained using the Percus–Yevick theory \cite{Percus-Yevick}, and $\Phi = \rho \pi d^3 / 6$ is the volume fraction. The first term on the right-hand side of Eq.~(\ref{eq:1}) represents the ideal fluid-like contribution, whereas the second term captures the caging constraints arising from interactions with neighboring particles.

\begin{figure*}[htp]
\includegraphics[width=18cm]{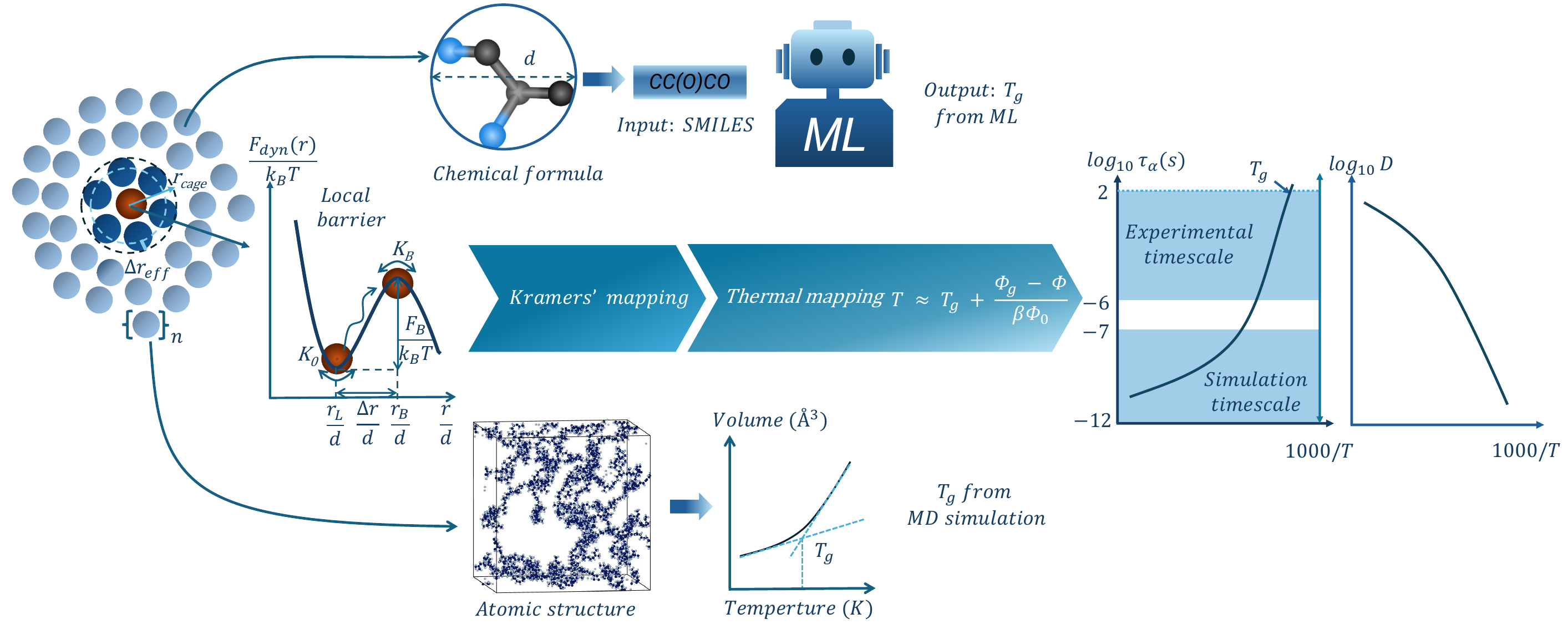}
\caption{\label{fig1}(Color online) Schematic illustration of the dynamic free energy in the ECNLE theory. Key length and energy quantities are defined.}
\end{figure*}

The dynamic free energy provides key insights into the local particle dynamics in dense fluids. At low volume fractions ($\Phi < 0.432$), the dynamic free energy $F_{\text{dyn}}(r)$ decreases monotonically with increasing $r$ and has no local barrier. In this regime, particles are not dynamically localized and diffuse freely due to weak interparticle interactions. This corresponds to fluid-like behavior \cite{15,16,18,19}. As the system becomes denser, a local barrier emerges in the dynamic free energy due to stronger interactions between the tagged particle and its nearest neighbors. These interactions give rise to transient cages that temporarily confine the particle's motion. The radius of the particle cage, $r_{\text{cage}}$, is determined as the position of the first minimum in the radial distribution function, $g(r)$. This function is related to the static structure factor $S(q)$ via an inverse Fourier transform. As density increases, the local barrier in the dynamic free energy becomes higher. The height of this barrier is given by $F_B = F_{\text{dyn}}(r_B) - F_{\text{dyn}}(r_L)$, where $r_L$ is the localization length and $r_B$ is the position of the barrier maximum in $F_{\text{dyn}}(r)$ as illustrated in Fig.~\ref{fig1}. The jump displacement $\Delta r = r_B - r_L$ represents the distance that the particle must move to escape its cage.

As a particle attempts to escape its cage, the surrounding particles must undergo cooperative rearrangements. This process leads to a slight expansion of the cage surface and generates a radially symmetric displacement field, $u(r)$, that propagates outward from the particle cage. Based on Lifshitz's continuous mechanics analysis \cite{Lifshitz}, the displacement field is 
\begin{eqnarray} 
u(r) = \Delta r_{eff}\frac{r_{cage}^2}{r^2},
\label{eq:3}
\end{eqnarray} 
where $\Delta r_{eff}$ is the amplitude of the cage expansion, which is
\begin{eqnarray} 
\Delta r_{eff} = \frac{3}{r_{cage}^3}\left[\frac{r_{cage}^2\Delta r^2}{32}- \frac{r_{cage}\Delta r^3}{192}+\frac{\Delta r^4}{3072}\right].
\label{eq:4}
\end{eqnarray}

The propagation of the displacement field induces vibrations in the surrounding particles. Since $u(r)$ is small, these vibrations can be approximated as harmonic and the elastic energy associated with a single particle is given by $\cfrac{1}{2}K_0 u^2(r)$ with $K_0 = \left|\partial^2 F_{dyn}(r)/\partial r^2\right|_{r=r_L}$ being the spring constant. To quantify the effect of collective motion on relaxation, the total elastic energy is obtained by summing the contributions from all particles outside the cage \cite{19,20,21,22,23,30,31,32,33,34,35,ECNLE1, ECNLE2, 36,37,38}
\begin{eqnarray}
F_e = 4\pi \rho \int_{r_{\text{cage}}}^{\infty} r^2 g(r) \frac{K_0 u^2(r)}{2} \, dr.
\label{eq:5}
\end{eqnarray}

Based on Kramers’ theory, we can calculate the structural relaxation time via \cite{19,20,21,22,23}
\begin{eqnarray} 
\frac{\tau_\alpha}{\tau_s} = 1 + \frac{2\pi}{\sqrt{K_0 K_B}} \frac{k_B T}{d^2} \exp\left(\frac{F_B + F_e}{k_B T}\right), 
\label{eq:6}
\end{eqnarray}
where $\tau_s$ is the short-time relaxation and $K_B = \left|\cfrac{\partial^2 F_{\text{dyn}}(r)}{\partial r^2}\right|_{r = r_B}$ is the magnitude of the curvature of the dynamic free energy at the peak of the barrier. The analytical expression for $\tau_s$ has been discussed in detail in prior studies \cite{ECNLE1, ECNLE2, 34, 35}.

Note that the calculations above provide $\tau_\alpha$ as a function of volume fraction $\Phi$. To quantitatively compare with experimental data, it is necessary to map $\Phi$ to temperature. To achieve this, Mirigian and Schweizer developed a thermodynamic mapping scheme for rigid molecular liquids~\cite{30,31,32}. This method relies on matching the dimensionless amplitude of long-wavelength density fluctuations, defined as $S_0^{\text{expt}}(T) = \rho k_B T \kappa_T$, where $\kappa_T$ is the isothermal compressibility. This quantity reflects nanoscale thermal density fluctuations and can be obtained from the experimental equation of state. In the hard-sphere fluid model with the Percus–Yevick approximation, the corresponding zero-wavenumber structure factor is given by $S_0^{\text{HS}} = \cfrac{(1 - \Phi)^4}{(1 + 2\Phi)^2}$. By equating $S_0^{\text{HS}}$ with $S_0^{\text{expt}}(T)$~\cite{19,20,21,22,23,30,31,32,33,34,35,ECNLE1, ECNLE2, 36,37,38}, we can determine the volume fraction as a function of temperature ($\Phi(T)$). This thermal mapping allows us to calculate the temperature dependence of $\tau_\alpha$ within the ECNLE theory.

Although ECNLE calculations associated with Schweizer's thermal mapping have been successful in predicting the dynamics of many glass-forming materials \cite{31, ECNLE1} without any adjustable parameter, its assumption of a universal link between local and long-range dynamics has a limitation. In reality, variations in microstructure and chemical interactions lead to significant differences in relaxation behavior, particularly in the dynamic fragility of polymers with different chemistries. To address this, Xie and Schweizer extended the ECNLE theory by introducing a non-universal coupling between local hopping and collective elastic barriers \cite{Xie}. They argued that the microscopic hopping distance needed for a particle to escape its cage is specific to the chemical structure and nanoscale rearrangements of the material. This chemical specificity is encoded into the theory through a single material-dependent parameter, $a_c$, which scales the collective elastic barrier. The new elastic barrier becomes $F_{e,\text{new}} = a_c F_e$, and the structural relaxation time in Eq. (\ref{eq:6}) is modified as \cite{19,20,21,22,23,30,31,32,33,34,35,ECNLE1, ECNLE2, 36,37,38}
\begin{equation}
\frac{\tau_{\alpha}}{\tau_{s}} = 1 + 
\frac{2\pi}{\sqrt{K_0 K_B}}\frac{k_B T}{d^2} 
\exp\left( \frac{F_B + a_c F_e}{k_B T} \right).
\label{eq:10}
\end{equation}
The extended theory significantly improves the predictions for the glass transition temperature, dynamic fragility, and temperature dependence of $\tau_\alpha$ simultaneously. However, calculations for the extended theory are highly dependent on equation-of-state data and the adjustable parameter $a_c$, which limits its practical application.

To address the limitations of previous thermal mapping schemes, we introduced an alternative approach based on the principle of thermal expansion
\begin{eqnarray} 
T \approx T_g + \frac{\Phi_g - \Phi}{\beta \Phi_0},
\label{eq:11}
\end{eqnarray}
where $T_g$ is the glass transition temperature defined at the structural relaxation time of 100 s, $\Phi_g$ is the corresponding volume fraction where $\tau_\alpha = 100$ s, $\Phi_0 = 0.5$ is a characteristic volume fraction, and $\beta \approx 12 \times 10^{-4}$ is the effective thermal expansion coefficient assumed constant for amorphous materials. Equation (\ref{eq:11}) shows that $\Phi$ can be directly mapped to temperature using only $T_g$ as input. Importantly, $\Phi_g$ depends on the material-specific parameter $a_c$ through Eq.~(\ref{eq:10}). For $a_c = 1$, the value of $\Phi_g$ is 0.6157, which is consistent with prior studies. This mapping significantly simplifies input requirements and is well suited for high-throughput application to diverse systems. The $T_g$ value can be obtained experimentally via DSC or BDS \cite{13, 16}, or alternatively estimated using computational simulations \cite{MD} or machine learning methods \cite{23, 36}.

The relationship between the diffusion coefficient ($D$) and the structural relaxation time is described by \cite{23,ECNLE3,ECNLE4}
\begin{eqnarray}
D(T)=\left(\frac{\Delta r}{d}\right)^2\frac{d^2}{6\tau_{\alpha} (T)}.
\label{eq:12}
\end{eqnarray}
In several previous studies \cite{23,ECNLE3,ECNLE4}, diffusion coefficients calculated using the ECNLE theory have shown good agreement with experimental measurements for various glass-forming systems \cite{23,ECNLE3,ECNLE4}.

{The ECNLE theory has several distinct advantages over other theoretical models such as mode-coupling theory (MCT) \cite{MCT1, MCT2}, random first order transition (RFOT) theory \cite{RFOT}, and the Adam-Gibbs (AG) model \cite{AG} to describe the thermal dependence of structural relaxation in supercooled liquids and polymers. The MCT captures short-time caging and two-step relaxation behavior and breaks down below the crossover temperature because it neglects activated hopping processes. The RFOT assumes the existence of cooperatively rearranging regions and a free-energy landscape without a precise microscopic formulation for the energy barrier. Similarly, although the AG model links relaxation time to configurational entropy, it remains largely phenomenological and lacks a clear microscopic basis. In contrast, ECNLE explicitly accounts for both local hopping out of transient cages and long-range collective elastic distortions. This dual contribution allows it to accurately model relaxation dynamics over a wide temperature range, including near and below the glass transition. In general, the ECNLE theory provides a more physical and predictive framework to understand relaxation time, fragility, and glass formation in a broad range of amorphous materials \cite{19,20,21,22,23,30,31,32,33,34,35,ECNLE1, ECNLE2, 36,37,38}.}

\subsection{Molecular dynamics simulation}
To estimate the glass transition temperature of PPS, PI, PPG, and PB, molecular dynamics simulations were conducted using the Large-scale Atomic/Molecular Massively Parallel Simulator (LAMMPS) \cite{LAMMPS}. {Although the MD cooling rates are significantly faster than those in experiments, our aim is not to exactly reproduce experimental $T_g$ values. Instead, we use the MD-derived $T_g$ values as system-specific inputs to ECNLE theory to model temperature-dependent dynamic properties without relying on experimental data. This allows us to build a self-contained predictive framework.} The calculation procedure involves three key steps: (1) constructing and minimizing initial configurations, (2) equilibrating and cooling the system under constant pressure and temperature, and (3) determining $T_g$ based on the simulated temperature dependence of the system volume.

%Despite the fast-cooling regime, our ECNLE calculations based on MD-derived $T_g$ values exhibit reasonable agreement with experimental data for structural relaxation time, diffusion constant, and dynamic fragility.
 
For each polymer, simulation cells containing either 10 or 20 chains with a polymerization degree of $N = 10$ were generated. These structures were packed into a periodic cubic box of dimensions $60 \times 60 \times 60$ \AA$^3$ to provide an initial mass density close to 1.0 g/cm$^3$. The Dreiding force field was employed due to its applicability to both organic molecules and polymeric systems \cite{force_field}. This force field accounts for bonded interactions (bond, angle, dihedral) and nonbonded interactions (van der Waals and hydrogen bonding), enabling realistic modeling of polymer conformations and intermolecular forces.

{The chosen polymerization degree of $N=10$ corresponds to molecular weights of approximately 631, 683, 656, and 570 g/mol for PPS, PI, PPG, and PB, respectively. While these values lie in the low-to-intermediate molecular weight regime, they are comparable to the experimental molecular weights of PI (1040 g/mol), PPG (192 g/mol), and PB (777 g/mol) \cite{25}. This allows for meaningful comparisons of their $T_g$ values and relaxation dynamics \cite{25}. Prior studies \cite{25} have shown that the glass transition temperature of PPG remains relatively insensitive across a broad molecular weight range (134–18000 g/mol), and the $T_g$ values of PB with molecular weights near 570 and 777 g/mol are nearly identical. Thus, for these polymers, although the molecular weight has significant effects on the glass transition temperature and fragility \cite{25, effect_M_T1, effect_M_T2}, our simulation conditions are representative of experimentally accessible behavior. However, for PPS, the simulated molecular weight is significantly lower than the experimental value ($\sim$ 44000 g/mol) \cite{25}, and quantitative comparisons of $T_g$ and fragility in this case are cautiously studied.}

Following energy minimization, each polymer system was equilibrated at a high temperature selected to ensure molecular mobility and thermal stability. The equilibration temperatures were set to 340 K for PPS, 300 K for PI and PPG, and 330 K for PB. Simulations were performed in the NPT ensemble using the Nosé–Hoover thermostat and barostat for 250 ps to maintain constant thermodynamic conditions. Subsequently, the systems were cooled in a stepwise manner to target temperatures of 150 K for PPS, 100 K for PI and PPG, and 50 K for PB. The cooling process was executed over multiple temperature intervals of 20–25 K with each interval involving 200 ps of equilibration. At each step, the specific volume was averaged over the final 100 ps of simulation. To determine $T_g$, the temperature–volume data were analyzed by fitting two linear regimes corresponding to high- and low-temperature regime. The intersection of these two linear fits was taken as the glass transition temperature for each polymer.

\subsection{Machine learning models}

In addition to MD simulation approach, machine learning can be exploited to predict the $T_g$ value of polymers. Among various ML techniques, Gaussian Process Regression (GPR) has recently been shown to provide highly accurate $T_g$ predictions for polymeric systems \cite{36}. Based on this, we selected a GPR model using a custom Tanimoto kernel to construct the predictive model in the present study. Since the experimental $T_g$ values of PPS, PI, PPG, and PB are all below 320 K, we selected a subset of 1586 polymers with experimental $T_g$ values under 320 K from a larger dataset comprising 7174 polymers with $T_g$ values ranging from 134 K to 768 K \cite{data}.  

The chemical structures of the polymers were first represented using the Simplified Molecular Input Line Entry System (SMILES) format \cite{smiles}, which encodes how atoms are connected within each molecule. These representations were then processed using RDKit \cite{RDKit} to generate quantitative molecular fingerprints that capture key features of the molecular topology, such as the arrangement of atoms and bonds. Each fingerprint consists of 2048 binary values (0 and 1), where each bit indicates the presence or absence of a specific structural pattern. These molecular descriptors were used as the input features for training and testing the machine learning model. All dataset and code used in this study are openly available on GitHub repository \cite{GPR_Model} to ensure full reproducibility and transparency. The dataset was randomly divided into training (80$\%$) and testing (20 $\%$) sets. To evaluate the predictive accuracy, we used two common performance metrics: the coefficient of determination ($R^2$), which indicates how well the model captures variance in the data, and the root mean square error (RMSE), which measures the average deviation between predicted and actual $T_g$ values.

\begin{figure*}[htp]
\includegraphics[width=18cm]{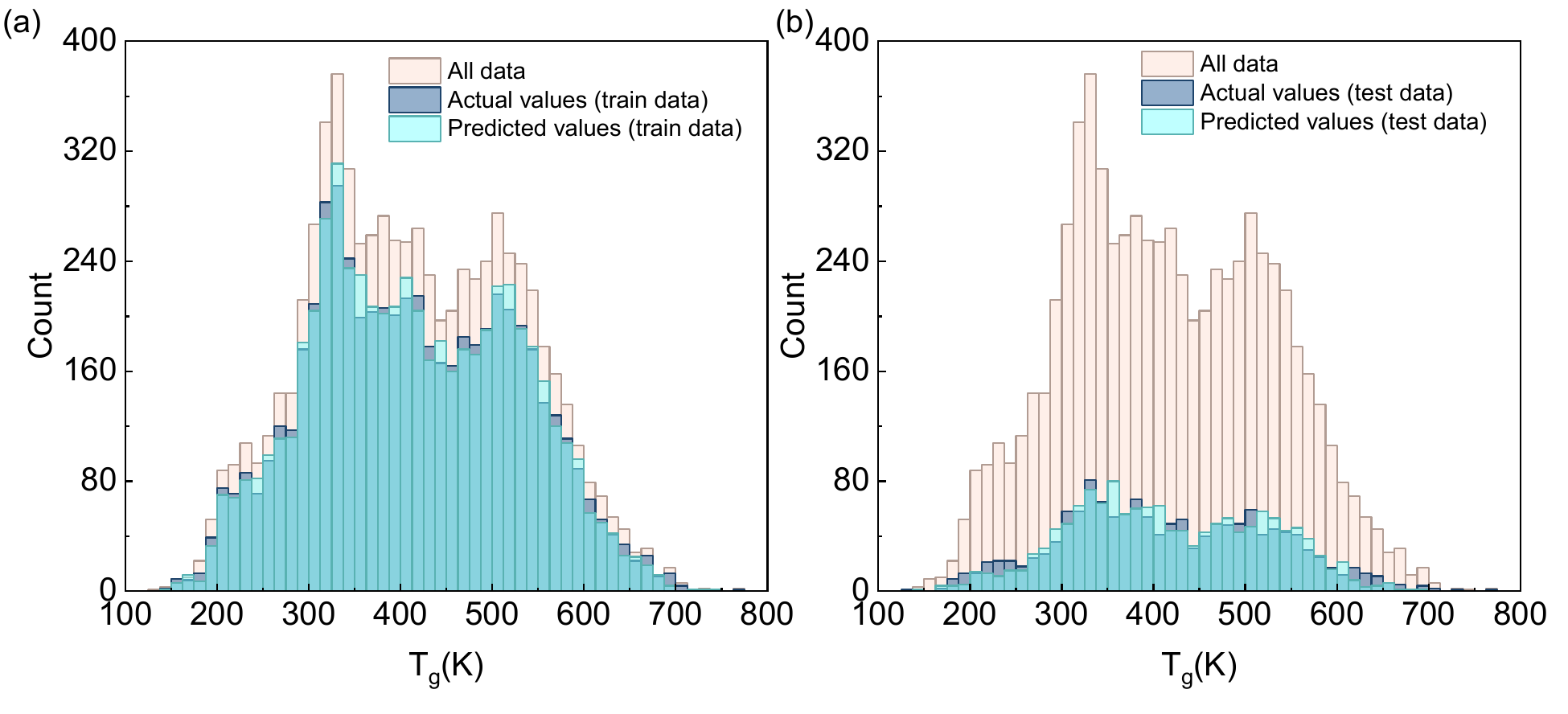}
\caption{\label{fig7}(Color online) {Distribution of actual and predicted glass transition temperatures of polymers for (a) training data and (b) test data.}}
\end{figure*}

{The histograms in Fig. \ref{fig7} show the frequency distributions of actual and predicted $T_g$ from our machine learning model. For the training set (Fig. \ref{fig7}a), the distribution of the predicted $T_g$ values shows excellent overlap and is nearly indistinguishable from the actual $T_g$ distribution. Crucially, for the test set (Fig. \ref{fig7}b), our predictions have a strong coherence between the actual and predicted distributions. Specifically, the predicted $T_g$ distribution in the test set successfully captures the overall shape, range, and modal features of the experimental data. This strong agreement in both training and test sets suggests that our model is reliable for predicting the $T_g$ values.}
\section{Results AND Discussion}
\begin{figure*}[htp]
\includegraphics[width=18cm]{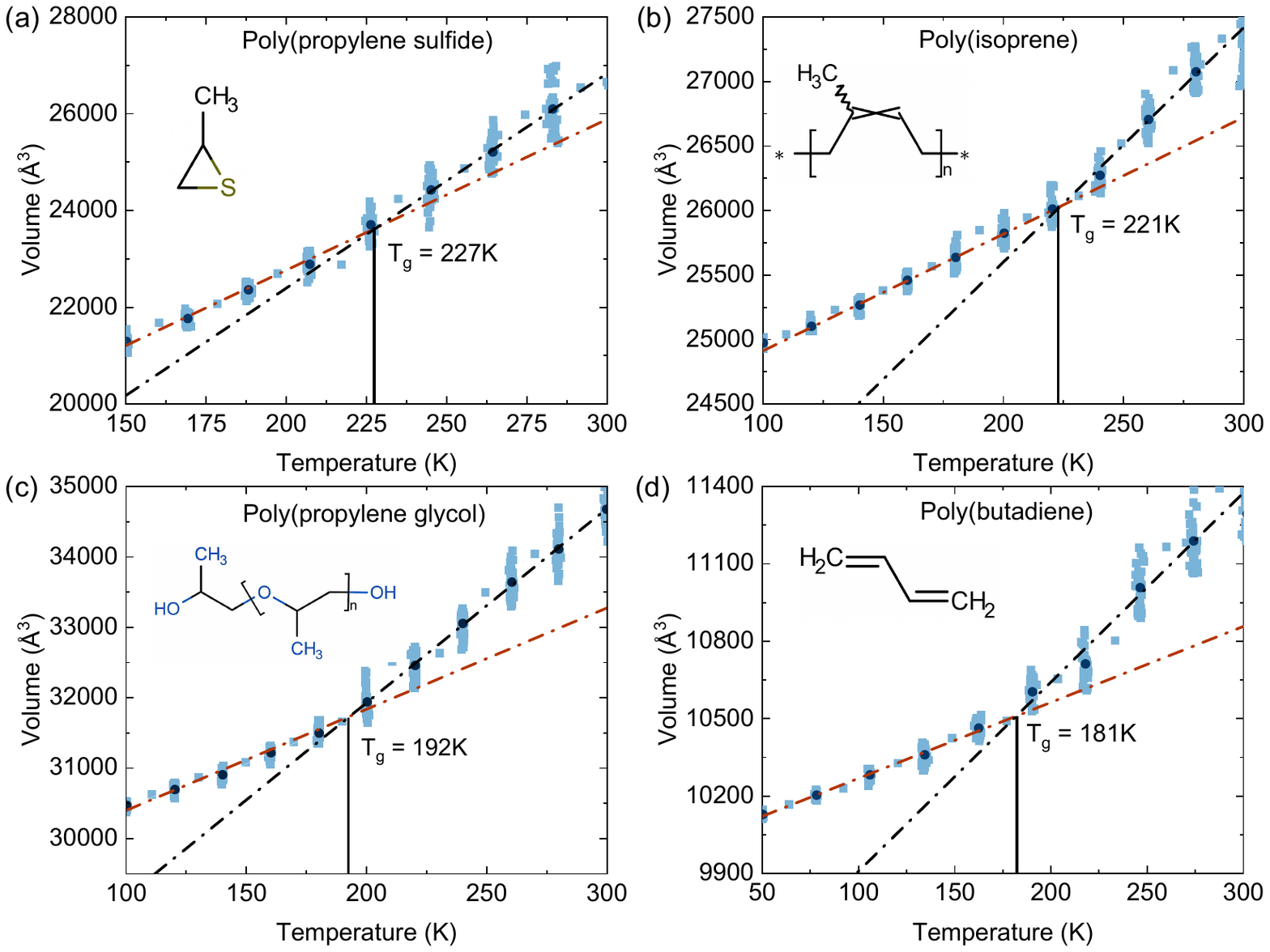}
\caption{\label{fig2}(Color online) Temperature dependence of the simulated volume from MD simulations for (a) PPS, (b) PI, (c) PPG, and (d) PB. Blue squares indicate raw data at each temperature step, while black circles indicate averaged values used for analysis. Linear fits to the low-temperature and high-temperature regimes are shown as brown and black lines, respectively. The intersection of these two lines corresponds to the predicted $T_g$ of the polymer.}
\end{figure*}

Figure~\ref{fig2} shows the temperature dependence of volume (in \AA$^{3}$) obtained from MD simulations for four polymer systems. Each system was equilibrated at a high temperature and subsequently cooled in discrete steps to a lower target temperature. This cooling process resulted in a gradual volume decrease. This change indicates the transition from a supercooled liquid with a higher thermal expansivity to a glassy state with significantly reduced molecular mobility. Linear fits were applied to the averaged volume data in both high- and low-temperature regions, and the intersection point of these lines was taken as the predicted $T_g$. The calculated $T_g$ values are 227~K for PPS, 221~K for PI, and 192~K for PPG.

\begin{table}[!ht]
    \centering
    \caption{\label{table1} The $T_g$ values of selected polymers obtained from MD simulations and ML calculations in this work, and prior experimental study \cite{25}.}
    \begin{tabular}{|c|c|c|c|}
    \hline
        Material & $T_{g,expt}$ (K) & $T_{g,MD}$ (K) & $T_{g,ML}$ (K) \\ \hline
        PPS & 229 & 227 & 249 \\ \hline
        PI & {185} & 221 & 240  \\ \hline
        PPG & {191} & 192 & 277  \\ \hline
        PB & {166} & 181 & 249   \\ \hline
    \end{tabular}
\end{table}

Our machine learning model, which is based on GPR, predicts $T_g$ with $R^2$ of 0.78 and RMSE of 19.5~K. {Figure~S2 in the Supplementary Information shows the predicted versus experimental $T_g$ values for the training and test sets of the low-$T_g$ subset to confirm comparable errors for both.} This performance indicates reasonable predictive accuracy within the low-$T_g$ range relevant to our selected polymers. Compared to a previous GPR study \cite{36}, which reported $R^2$ of over 0.89 and an RMSE below 37~K for a full dataset spanning $T_g$ values from 134 to 768~K, our model has a lower overall $R^2$ but a smaller RMSE. {This difference arises because we focus our training on a subset of 1586 polymers with $T_g$ values below 320~K to better represent the temperature range of interest. As detailed in Tables~SI–SIII of the Supplementary Information, cross-validation shows that this choice yields slightly lower global $R^2$ due to the narrower $T_g$ range, but improves accuracy (lower RMSE) for low-$T_g$ polymers and avoids the systematic upward shift in predicted $T_g$ that occurs when higher-$T_g$ polymers are included in the training set.}

{Several recent studies have applied more complex machine-learning architectures to predict polymer $T_g$. A graph neural network (GNN) model was specifically developed for polyimides, focusing on high-$T_g$ materials within a relatively narrow chemical space \cite{40}. However, its maximum $R^2$ of 0.86 implies a non-negligible amount of unexplained variance and does not provide higher accuracy than our GPR model. Park \emph{et al.}\ combined a graph convolutional network (GCN) with a neural-network (NN) regressor to learn $T_g$ and other properties over a dataset with a similar $T_g$ range and distribution to our full database \cite{41}. Although their deep-learning architecture is more complex, the resulting prediction accuracy is only comparable to that of our simpler GPR approach. Cheminformatics models coupled to coarse-grained MD simulations \cite{42} can identify physical descriptors, but these simulations are time-consuming and do not always reproduce experimental behavior accurately. As a result, this approach yields only $0.47 \le R^2 \le 0.74$ for $T_g$ prediction across diverse polymers and is difficult to deploy in a high-throughput setting. More recently, quantum-chemical-augmented NN/GNN frameworks \cite{43} that rely on quantum-chemistry-derived local-cluster descriptors report $R^2 \approx 0.74$ for $T_g$ and require additional quantum-chemical calculations for each polymer. In contrast, our GPR model operates on simple binary fingerprints generated from SMILES strings, avoids any MD- or QC-based feature generation, and still achieves higher accuracy in the low-$T_g$ regime of interest. This balance of accuracy, simplicity, and computational efficiency makes the GPR framework the most appropriate choice for the present work.}

To further evaluate the accuracy of both ML and MD approaches, we compare their predicted $T_g$ values with experimental data in Ref.~\cite{25}. As summarized in Table~\ref{table1}, the MD results show good quantitative agreement with experimental values and deviations are typically within 10–15~K. The GPR model predicts the $T_g$ values for PPS and PI with moderate accuracy. The ML predictions overestimate the experimental counterparts by 20~K and 36~K, respectively. {For PPG and PB, the deviations are larger, on the order of 80~K. This likely indicates both the limited number of structurally similar low-$T_g$ polymers in the training data and the intrinsic scatter of the experimental $T_g$ database. Hence, in this work we use the GPR model mainly to capture overall trends in the low-$T_g$ regime and to provide an additional input for testing ECNLE predictions, rather than as a highly accurate predictor for each individual flexible polymer.} While MD simulation provides more accurate absolute $T_g$ predictions, the GPR model remains a useful and efficient tool for trend prediction and rapid screening, especially when experimental data are unavailable. {A systematic exploration of more expressive architectures and additional descriptors as shown in Ref. \cite{40,41,42,43} to further reduce these errors is an important direction for future work, but is beyond the scope of this study, which focuses on testing ECNLE predictions against different $T_g$ inputs.}

\begin{figure*}[htp]
\includegraphics[width=18cm]{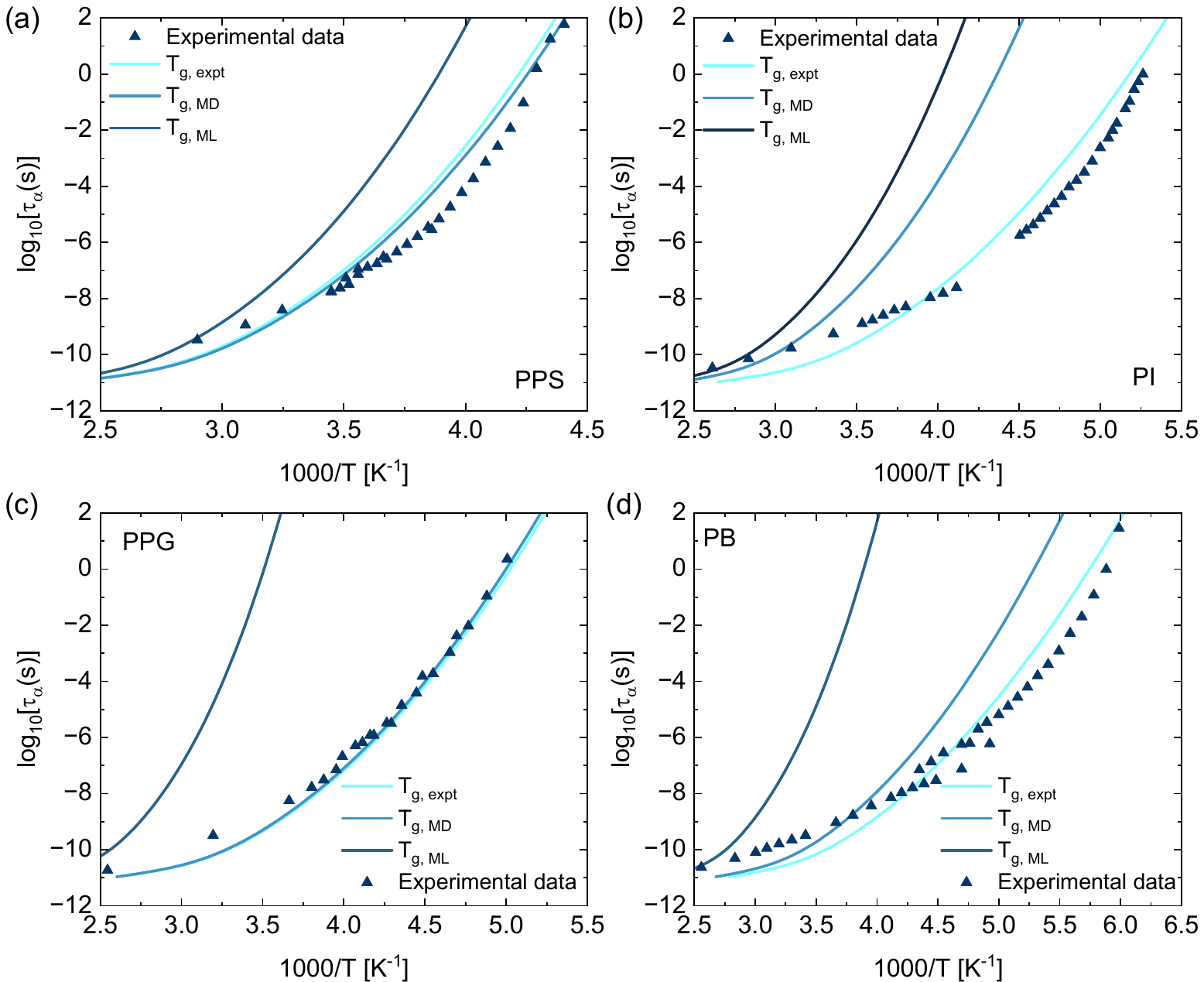}
\caption{\label{fig3}(Color online) Temperature dependence of the structural relaxation time for (a) PPS, (b) PI, (c) PPG, and (d) PB calculated using Eqs. (\ref{eq:1}-\ref{eq:6}) and (\ref{eq:11}). The $T_g$ used in the thermal mapping (Eq.~(\ref{eq:11})) is obtained from MD simulations, ML predictions, and experimental data in Ref. \cite{25}. {Experimentally, the molecular weights obtained were 44000 g/mol for PPS, 1040 g/mol for PI, 192 g/mol for PPG, and 777 g/mol for PB. In contrast, the corresponding values derived from MD simulation were 631, 683, 656, and 570 g/mol, respectively.}}
\end{figure*}

Figure \ref{fig3} shows the structural relaxation time as a function of $1000/T$ for PPS, PI, PPG, and PB. The relaxation times were first calculated as a function of volume fraction using Eqs. (\ref{eq:1})-(\ref{eq:6}), and then converted to temperature using the thermal mapping described in Eq. (\ref{eq:11}), where $T_g$ is a key input. The $T_g$ values used in this mapping were obtained from MD simulations, ML predictions, and experimental data as listed in Table~\ref{table1}. ECNLE predictions using experimental $T_g$ values ($T_{g,\text{expt}}$) quantitatively agree with BDS data in Ref.~\cite{25} for all polymers. When using MD-predicted $T_g$ values ($T_{g,\text{MD}}$), the predicted relaxation times remain consistent with experimental data for most systems, although a noticeable deviation is observed for PI. This finding is predictable since the difference between $T_{g,\text{MD}}$ and $T_{g,\text{expt}}$ is largest for PI. In contrast, using ML-predicted $T_g$ values ($T_{g,\text{ML}}$) leads to greater discrepancies between ECNLE results and experimental data, particularly at lower temperatures.

\begin{figure*}[htp]
\includegraphics[width=18cm]{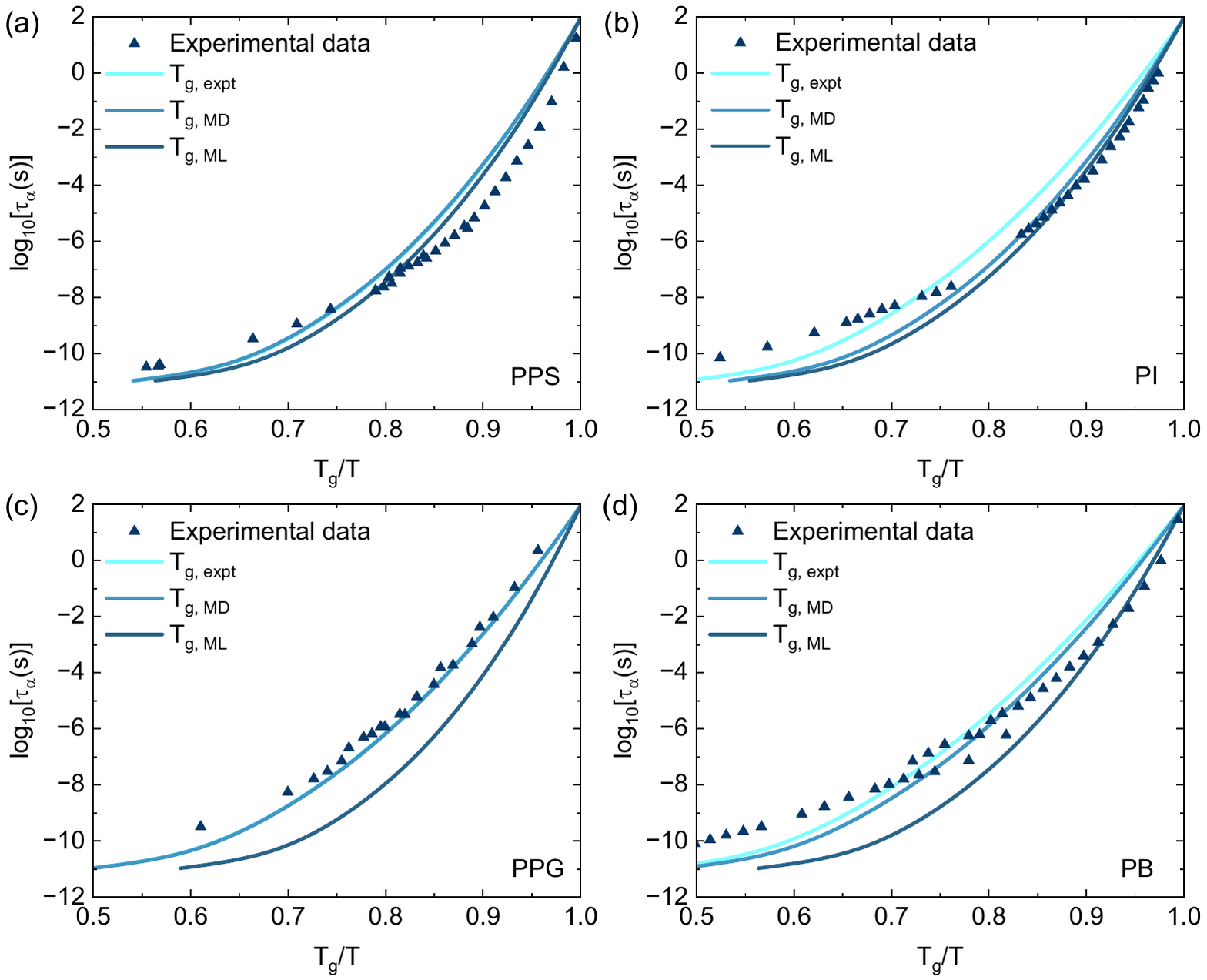}
\caption{\label{fig4}(Color online) The logarithm of structural relaxation times as a function of $T_g/T$ for (a) PPS, (b) PI, (c) PPG, and (d) PB from the same data as in Fig. \ref{fig3}.}
\end{figure*}

To facilitate direct comparison with experimental results, the relaxation time data from Fig.~\ref{fig3} are replotted as a function of $T_g/T$ in Fig.~\ref{fig4}. This normalized representation shows that the results from the MD simulations and the GPR model agree more closely with the experimental data than when plotted in Fig.~\ref{fig3}. Moreover, plotting $\log \tau_\alpha$ versus $T_g/T$ allows us to calculate the dynamic fragility, which is defined as $m = \left[ \cfrac{d \log \tau_\alpha}{d (T_g / T)} \right]_{T = T_g}$~\cite{39}. This quantity quantifies the steepness of the increase in the structural relaxation time near $T_g$. Based on the value of the dynamic fragility index $m$, glass-forming materials are typically categorized into three groups. Materials with $m < 30$ are considered "strong" glass formers and their $\tau_\alpha(T)$ exhibits relatively Arrhenius-like behavior. Those with $m > 100$ are classified as "fragile" and the temperature dependence of the structural relaxation time shows highly non-Arrhenius dynamics near $T_g$. Systems with $m$ values between 30 and 100 are regarded intermediate glass formers.

\begin{table}[!ht]
    \centering
    \caption{ The dynamic fragility values of selected polymers computed using ECNLE theory with $T_g$ inputs from experiments, MD simulations, and ML predictions as listed in Table~\ref{table1}, compared with experimental fragility.}
    \begin{tabular}{|c|c|c|c|c|c|}
    \hline
        Material  &  $m_{\text{expt}}$ & $m$ ($T_{g,expt}$) & $m$ ($T_{g,MD}$) & $m$ ($T_{g,ML}$) \\ \hline
        PPS  & {116} & 84 & 84 & 89 \\ \hline
        PI & {80} & 75.4 & 82 & 89   \\ \hline
        PPG  &{79} & 67 & 67 &  103  \\ \hline
        PB  & {84} & 64 & 67 & 92 \\ \hline
    \end{tabular}
    \label{table2}
\end{table}

Table~\ref{table2} summarizes the dynamic fragility calculated using ECNLE theory using different $T_g$ inputs and experimental fragility values. In most cases, the calculated $m$ values are lower than the experimental values, but the deviations remain within an acceptable range. When using experimental $T_g$ as input, the calculated $m$ values for PPS and PI are within 10–15 $\%$ of experimental fragilities. Although the $T_{g,MD}$ values are close to the experimental $T_g$, the fragilities calculated using $T_{g,ML}$ actually show better agreement with experimental values across all polymers. This suggests that machine learning, even with slightly overestimated $T_g$ values, can effectively predict the dynamic fragility. The observed differences may result from experimental uncertainties, the sensitivity of glassy dynamics to small variations in $T_g$, and approximations inherent in the theoretical model. {We also train our ML model based on the full dataset of polymer and find that the calculated fragilities deviate more significantly from experimental values (see in Supplementary Information). This further validates the importance of appropriate data selection.} Overall, these findings reveal that combining ECNLE theory with $T_g$ inputs from either MD or ML provides a reliable and practical approach for predicting the thermal behavior of polymer glasses.

To examine how specific molecular characteristics such as intramolecular flexibility, side-chain dynamics, and intermolecular interactions affect the $\tau_\alpha(T)$ calculations, we adopted the method proposed by Xie and Schweizer \cite{Xie}. This approach uses a material-specific parameter $a_c$ to adjust the contribution of the collective elastic barrier to the total energy barrier ($F_B +a_cF_e$). All above calculations in this work use the default value $a_c=1$. Figure~\ref{fig3} presents the structural relaxation times of PPS, PI, PPG, and PB as a function of $1000/T$ computed using the ECNLE theory with varying values of the parameter $a_c$ based on Eqs.~(\ref{eq:1})–(\ref{eq:5}), (\ref{eq:10}), and (\ref{eq:11}). Adjusting $a_c$ improves quantitative agreement between numerical results and experimental data. {This finding indicates that accurate prediction of $T_g$ alone is insufficient to quantitatively capture the temperature dependence of structural relaxation time, as the dynamic behavior is also sensitive to how the collective elastic barrier is treated.} However, even with optimal tuning, the model does not fully reproduce the relaxation behavior over the entire temperature range. It is also important to note that the predictions of this extended approach strongly depends on the selection of $a_c$, which currently lacks a first-principles basis and must be empirically adjusted. The dependence of this method on fitting limits its use to systems for which experimental data already exist.

\begin{figure*}[htp]
\includegraphics[width=18cm]{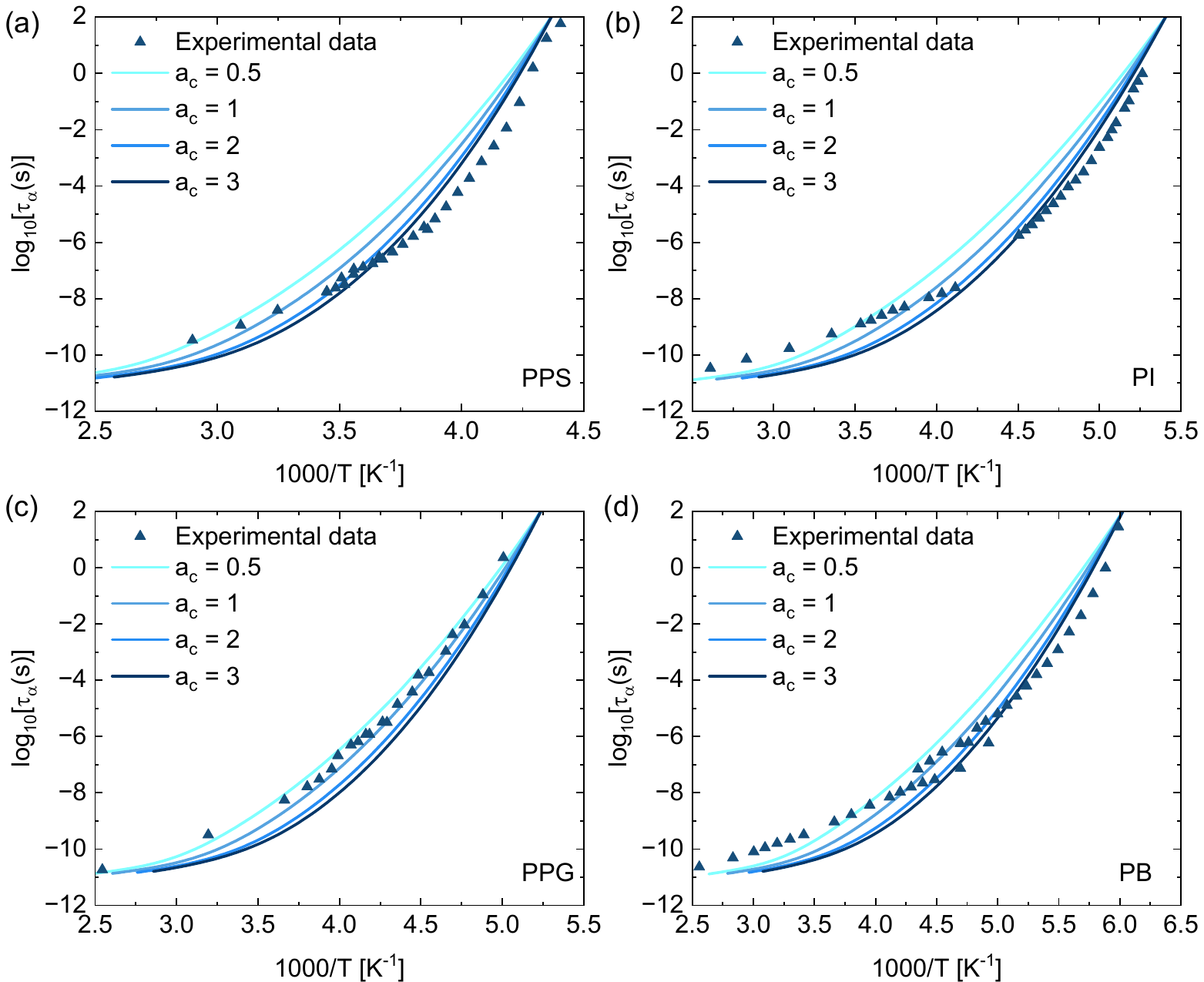}
\caption{\label{fig5}(Color online) Temperature dependence of the structural relaxation time for (a) PPG, (b) PI, (c) PPS, and (d) PB. Data points correspond to experimental data in Ref. \cite{25}. Solid curves are ECNLE theoretical predictions computed using Eqs. (\ref{eq:1}–\ref{eq:5}), (\ref{eq:10}), and (\ref{eq:11}) with experimental $T_g$ values and  different $a_c$ parameters.}
\end{figure*}

\begin{figure*}[htp]
\includegraphics[width=18cm]{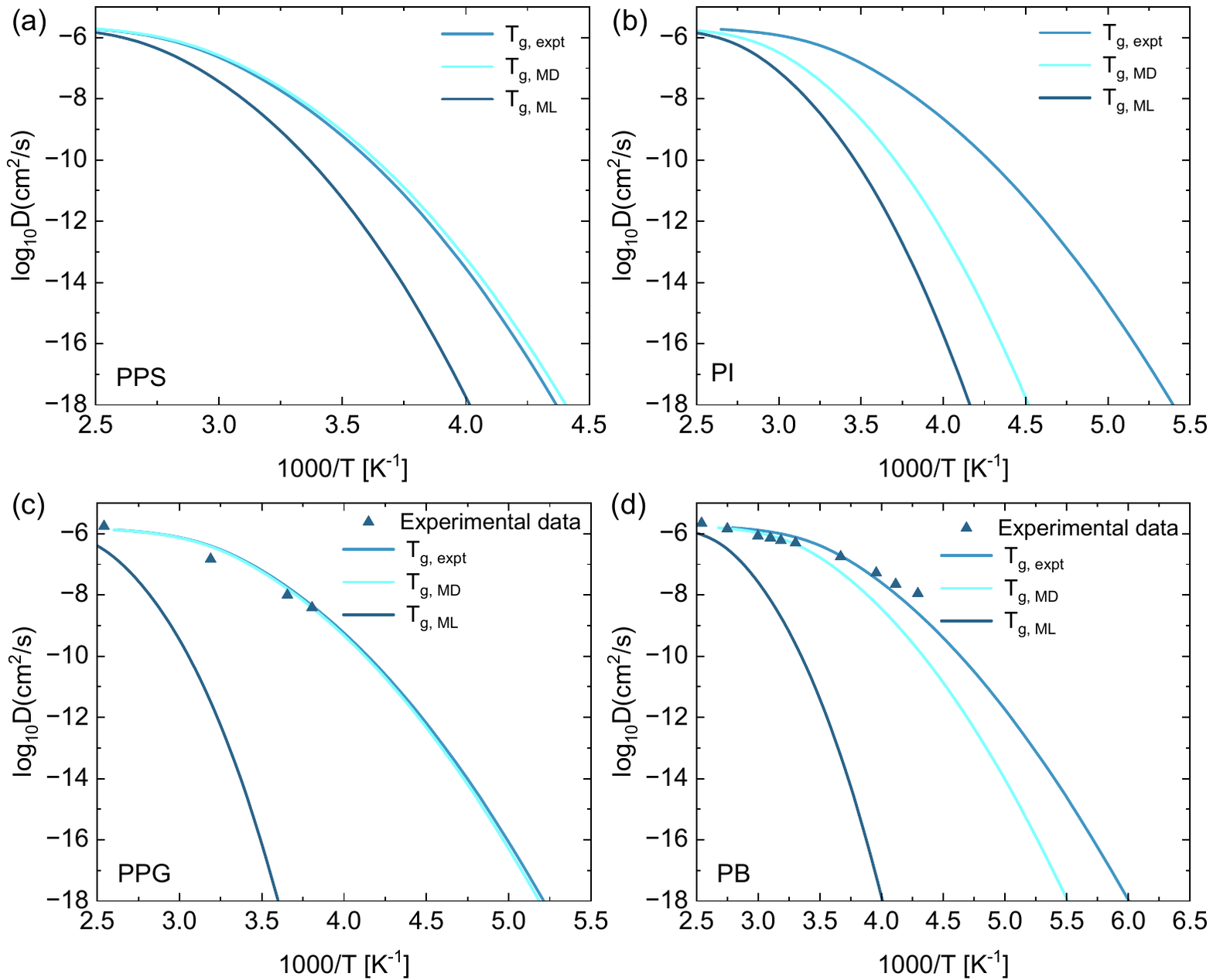}
\caption{\label{fig6}(Color online) The predicted temperature dependence of the diffusion coefficient for (a) PPS, (b) PI, (c) PPG, and (d) PB calculated using Eqs. (\ref{eq:1}-\ref{eq:6}), (\ref{eq:11}), and (\ref{eq:12}). The $T_g$ used in the thermal mapping (Eq.~(\ref{eq:11})) is obtained from experimental, MD simulations, ML predictions, {and experimental data \cite{diffusion}.}  {The molecular weights for PPG were 770 g/mol experimentally and 656 g/mol via MD simulation, while the values for PB were 466 g/mol and 570 g/mol, respectively.}}
\end{figure*}

The above results indicate that using $a_c$ = 1 in the ECNLE theory provides a reasonable prediction of $\tau_\alpha(T)$. Based on this, we apply Eq. (\ref{eq:12}) to compute the diffusion coefficient as a function of $T_g/T$ for PPS, PI, PPG, and PB. The particle diameter $d$, which is used to calculate the diffusion coefficient, was estimated for each polymer using the atomic structure of its monomer derived from SMILES codes and processed with the RDKit software. The computed diffusion coefficients are presented in Figure~\ref{fig6}. Using the $T_g$ values obtained from MD, ML, and experimental data, we find that the diffusion coefficient decreases rapidly as temperature decreases. Notably, the ECNLE predictions using $T_{g,MD}$ and $T_{g,expt}$ are relatively close, despite the quantitative differences between these $T_g$ values. Although the ECNLE theory is a microscopic statistical mechanical theory and does not explicitly account for spatially heterogeneous dynamics or the decoupling behavior between diffusion and relaxation time observed near the glass transition temperature \cite{add1, add2, add3}, our calculated diffusion coefficients show good agreement with experimental data for PPG and PB in Ref. \cite{diffusion}. This agreement reinforces the physical relevance of our predictions and suggests that, for the systems studied here, the ECNLE theory can provide a quantitatively accurate description of average translational dynamics.

\section{Conclusion}
In conclusion, we have developed a multiscale approach combining MD simulations, machine learning, and the ECNLE theory to study glass transition and relaxation dynamics in polymers. The glass transition temperatures of four representative polymers were estimated using both MD simulations and the GPR model trained on experimental data. These values were then used to calculate temperature-dependent structural relaxation times, diffusion coefficients, and dynamic fragility using the ECNLE theory. The MD predicted values $T_g$ agree well with the experimental data within 10–15 K. Meanwhile, the $T_g$ predictions based on machine learning are relatively higher than the experimental $T_g$s. ECNLE calculations using the values $T_g$ from the MD simulations and experiments show good agreement with the BDS data when plotted as $\log \tau_\alpha$ versus $1000/T$. In the Angell representation ($\log \tau_\alpha$ versus $T_g/T$), the ECNLE results based on $T_{g,MD}$, $T_{g,ML}$ and $T_{g,expt}$ are all quantitatively in agreement with the experimental relaxation data. This representation can reliably estimate the dynamic fragility. In particular, the fragilities calculated using $T_{g,ML}$ show better agreement with the experimental values compared to those based on $T_{g,MD}$. This suggests that machine learning provides not only trend preservation but also quantitative accuracy in fragility prediction. Although the extended ECNLE model allows tuning through a material-specific parameter $a_c$ to scale effects of collective dynamics on the glass transition, we found that the standard formulation ($a_c = 1$) already provides reasonable accuracy without any adjustable parameter. Overall, our study clearly validates the predictivity of our integrated approach, especially when experimental data are limited.

\begin{acknowledgments}
This research was funded by the Vietnam National Foundation for Science and Technology Development (NAFOSTED) under Grant No.103.01-2023.62.
\end{acknowledgments}

\section*{Conflicts of interest}
The authors have no conflicts to disclose.

\section*{Data availability}

% The data that support the findings of this study are available from the corresponding author upon reasonable request.
%No primary research results, software or code have been included and no new data were generated or analysed as part of this review.

Data and code for this article are available at GitHub at \url{https://github.com/NgoQue/GPR_Model}

\end{document}